\documentclass[10pt,twocolumn,letterpaper]{article}

\usepackage{common/cvpr}
\usepackage{times}
\usepackage{epsfig}
\usepackage{graphicx}
\usepackage{amsmath}
\usepackage{amssymb}
\usepackage[normalem]{ulem}
\useunder{\uline}{\ul}{}

\usepackage[pagebackref=true,breaklinks=true,colorlinks,bookmarks=false]{hyperref}

\cvprfinalcopy

\ifcvprfinal\fi
\begin{document}

\title{Mastering Terra Mystica: Applying Self-Play to Multi-agent Cooperative Board Games}

\author{
Luis Perez\\
Stanford University\\
450 Serra Mall\\
{\tt\small luisperez@cs.stanford.edu}
}

\maketitle
\thispagestyle{empty}

\begin{abstract}
In this paper, we explore and compare multiple algorithms for solving the complex strategy game of Terra Mystica, thereafter abbreviated as TM. Previous work in the area of super-human game-play using AI has proven effective, with recent break-through for generic algorithms in games such as Go, Chess, and Shogi \cite{AlphaZero}. We directly apply these breakthroughs to a novel state-representation of TM with the goal of creating an AI that will rival human players. Specifically, we present initial results of applying AlphaZero to this state-representation, and analyze the strategies developed. A brief analysis is presented. We call this modified algorithm with our novel state-representation AlphaTM. In the end, we discuss the success and short-comings of this method by comparing against multiple baselines and typical human scores. All code used for this paper is available at on \href{https://github.com/kandluis/terrazero}{GitHub}.

\end{abstract}

\section{Background and Overview}
\label{section:background_and_overview}
\subsection{Task Definition}
\label{section:task_definition}
In this paper, we provide an overview of the infrastructure, framework, and models required to achieve super-human level game-play in the game of Terra Mystica (TM) \cite{TMBoardGeek}, without any of its expansions \footnote{There are multiple expansions, most which consists of adding different Factions to the game or extending the TM Game Map. We hope to have the time to explore handling these expansions, but do not make this part of our goal}. The game of TM involves very-little luck, and is entirely based on strategy (similar to Chess, Go, and other games which have recently broken to novel Reinforcement Learning such Deep Q-Learning and Monte-Carlo Tree Search as a form of Policy Improvement \cite{AlphaGo} \cite{AlphaGoZero}). In fact, the only randomness arises from pregame set-up, such as players selecting different Factions \footnote{A Faction essentially restricts each Player to particular areas of the map, as well as to special actions and cost functions for building}, different set of end-round bonus tiles being selected.

TM is a game played between 2-5 players. For our research, we focus mostly on the adversarial 2-player version of the game. We do this mostly for computational efficiency, though for each algorithm we present, we discuss briefly strategies for generalizing them to multiple players. We also do this so we can stage the game as a zero-sum two-player game, where players are rewarded for winning/loosing.

TM is a fully-deterministic game whose complexity arises from the large branching factor and a large number of possible actions $A_t$ from a given state, $S_t$. There is further complexity caused by the way in which actions can interact, discussed further in Section \ref{subsection:input/output_of_system}.
\subsection{Input/Output Of System}
\label{subsection:input/output_of_system}
In order to understand the inputs and outputs of our system, the game of TM must be fully understood. We lay out the important aspects of a given state below, according to the standard rules \cite{TMRules}.

The game of TM consists of a terrain board that's split into $9 \times 13$ terrain tiles. The board is fixed, but each terrain tile can be terra-formed (changed) into any of the 7 distinct terrains (plus water, which cannot be modified). Players can only expand onto terrain which belongs to them. Furthermore, TM also has a mini-game which consists of a Cult-Board where individual players can choose to move up in each of the cult-tracks throughout game-play. 

The initial state of the game consists of players selecting initial starting positions for their original dwellings. 

At each time-step, the player has a certain amount of resources, consisting of Workers, Priests, Power, and Coin. The player also has an associated number of VPs.

Throughout the game, the goal of each player is to accumulate as many VPs as possible. The player with the highest number of VPs at the end of the game is the winner.

From the definition above, the main emphasis of our system is the task of taking a single state representation $S_t$ at a particular time-step, and outputting an action to take to continue game-play for the current player. As such, the input of our system consists of the following information, fully representative of the state of the game:

\begin{enumerate}
    \item Current terrain configuration. The terrain configuration consists of multiple pieces of information. For each terrain tile, we will receive as input:
    \begin{enumerate}
        \item The current color of the tile. This gives us information not only about which player currently controls the terrain, but also which terrains can be expanded into.
        \item The current level of development for the terrain. For the state of development, we note that each terrain tile can be one of (1) UNDEVELOPED, (2) DWELLING, (3) TRADING\_POST, (4) SANCTUARY, or (5) STRONGHOLD.
        \item The current end-of-round bonus as well as future end-of-round bonus tiles.
        \item Which special actions are currently available for use.
    \end{enumerate}
    \item For each player, we also receive the following information.
    \begin{enumerate}
        \item Current level of shipping ability, Current level of spade ability, the current number of VPs that the player has.
        \item The current number of towns the player has (as well as which town is owned), The current number of worker available to the player, the current number of coins available to the player, the current number of LV1, LV2, and LV3 power tokens.
        \item The current number of priests available to the player.
        \item Which bonus tiles are currently owned by the player.
        \item The amount of income the player currently produces. This is simply the power, coins, priests, and worker income for the player.
    \end{enumerate}
\end{enumerate}

The above is a brief summary of the input to our algorithm. However, in general, the input to the algorithm is a complete definition of the game state at a particular turn. Note that Terra Mystica \textit{does not} have any dependencies in previous moves, and is completely Markovian. As such, modeling the game as an MDP is fully realizable, and is simply a question of incorporating all the features of the state.

\subsection{Output and Evaluation Metrics}
For a given state, the output of our algorithm will consist of an action which the player can take to continue to the next state of the game. Actions in TM are quite varied, and we do not fully enumerate them here. In general, however, there are eight possible actions:
\begin{enumerate}
    \item Convert and build.
    \item Advance on the shipping ability.
    \item Advance on the spade ability.
    \item Upgrade a building.
    \item Sacrifice a priest and move up on the cult track.
    \item Claim a special action from the board.
    \item Some other special ability (varies by class)
    \item Pass and end the current turn.
\end{enumerate}

We will evaluate our agents using the standard simulator. The main metric for evaluation will be the maximum score achieved by our agent in self-play when winning, as well as the maximum score achieved against a set of human competitors.

\section{Experimental Approach}
\label{section:experimental_approach}
Developing an AI agent that can play well will be extremely challenging. Even current heuristic-based agents have difficulty scoring positions. The state-action space for TM is extremely large. Games typically have trees that are $>50$ moves deep (per player) and which have a branching factor of $>10$. 

We can approach this game as a typical min-max search-problem. Simple approaches would simply be depth-limited alpha-beta pruning similar to what we used in PacMan. These approaches can be tweaked for efficiency, and are essentially what the current AIs use.

Further improvement can be made to these approaches by attempting to improve on the $Eval$ functions.

However, the main contribution of this paper will be to apply more novel approaches to a custom state-space representation of the game. In fact, we will be attempting to apply Q-Learning -- specifically DQNs (as per \cite{AlphaGo}, \cite{AlphaGoZero}, and \cite{AlphaZero}). 

\subsection{Deep Q-Learning Methods for Terra Mystica}
\label{section:deep_qlearning_methods_for_tm}
Existing open-source AIs for TM are based on  combination of sophisticated search techniques (such as depth-limited, alpha-beta search, domain-specific adaptations, and handcrafted evaluation functions refined by expert human players). Most of these AIs fail to play at a competitive level against human players. The space of open-source AIs is relatively small, mainly due to the newness of TM.

\subsection{Alpha Zero for TM}
\label{subsection:alpha_zero_for_tm}
In this section, we describe the main methods we use for training our agent. In particular, we place heavy emphasis on the methods described by AlphaGo\cite{AlphaGo}, AlphaGoZero, \cite{AlphaGoZero}, and AlphaZero \cite{AlphaZero} with pertinent modifications made for our specific problem domain.

Our main method will be a modification of the Alpha Zero \cite{AlphaZero} algorithm which was described in detail. We chose this algorithm over the methods described for Alpha Go \cite{AlphaGo} for two main reasons:
\begin{enumerate}
    \item The Alpha Zero algorithm is a zero-knowledge reinforcement learning algorithm. This is well-suited for our purposes, given that we can perfectly simulate game-play. 
    \item The Alpha Zero algorithm is a simplification over the dual-network architecture used for AlphaGo.
\end{enumerate}

As such, our goal is to demonstrate and develop a slightly modified general-purpose reinforcement learning algorithm which can achieve super-human performance tabula-rasa on TM.

\subsubsection{TM Specific Challenges}
\label{subsubsection:tm_specific_challenges}
We first introduce some of the TM-specific challenges our algorithm must overcome.

\begin{enumerate}
    \item Unlike the game of Go, the rules of TM are \textbf{not} translationally invariant. The rules of TM are position-dependent -- the most obvious way of seeing this is that each terrain-tile and patterns of terrains are different, making certain actions impossible from certain positions (or extremely costly). This is not particularly well-suited for the weight-sharing structure of Convolutional Neural Networks.
    \item Unlike the game of Go, the rules for TM are asymmetric. We can, again, trivially see this by noting that the board game board itself \ref{fig:TM_Board} has little symmetry.
    \item The game-board is not easily quantized to exploit positional advantages. Unlike games where the AlphaZero algorithm has been previously applied (such as Go/Shogi/Chess), the TM map is not rectangular. In fact, each ``position'' has \textbf{6} neighbors, which is not easily representable in matrix form for CNNs.
    \item The action space is significantly more complex and hierarchical, with multiple possible ``mini''-games being played. Unlike other games where similar approaches have been applied, this action-space is extremely complex. To see this, we detail the action-spaces for other games below.
    \begin{enumerate}
        \item The initial DQN approach for Atari games had an output action space of dimension 18 (though some games had only 4 possible actions, the maximum number of actions was 18 and this was represented simply as a 18-dimensional vector representing a softmax
        probability distribution).
        \item For Go, the output actions space was similarly a $19\times19 + 1$ probability distribution over the locations on which to place a stone. 
        \item Even for Chess and Shogi, the action space similarly consisted of all legal destinations of all player's pieces on the board. While this is very expansive and more similar to what we expect for TM, TM nonetheless has additional complexity in that some actions are inherently hierarchical. You must first decide if you want to build, then decided \textbf{where} to build, and finally decide \textbf{what} to build. This involves defining an output actions-space which is significantly more complex than anything we've seen in the literature. For comparison, in \cite{AlphaZero} the output space consists of a stack of planes of $8\times 8 \times 73$. Each of the 64 positions identifies a piece to be moved, with the 73 associated layers identifying exactly \textit{how} the piece will be moved. As can be seen, this is essentially a two-level decision tree (select piece followed by selecting how to move the piece). In TM, the action-space is far more varied.
    \end{enumerate}
    \item The next challenge is that TM is not a binary win-lose situation, as is the case in Go. Instead, we must seek to maximize our score \textit{relative} to other players. Additionally, in TM, there is always the possibility of a tie. 
    \item Another challenge present in TM not present in other stated games is the fact that there exist a limited number of resources in the game. Each player has a limited number of workers/priests/coin with which a sequence of actions must be selected.
    \item Furthermore, TM is now a multi-player game (not two-player). For our purposes, however, we leave exploring this problem to later research. We focus exclusively on a game between two fixed factions (Engineers and Halflings).
\end{enumerate}

\subsection{Input Representation}
\label{subsection:input_representation}
Unless otherwise specified, we leave the training and search algorithm large unmodified from those presented in \cite{AlphaZero} and \cite{AlphaGoZero}. We will described the algorithm, nonetheless, in detail in subsequent sections. For now, we focus on presented the input representation of our game state.

\subsubsection{The Game Board}
\label{subsubsection:the_game_board}

\begin{figure}[!ht]
    \centering
    \includegraphics[scale=0.2]{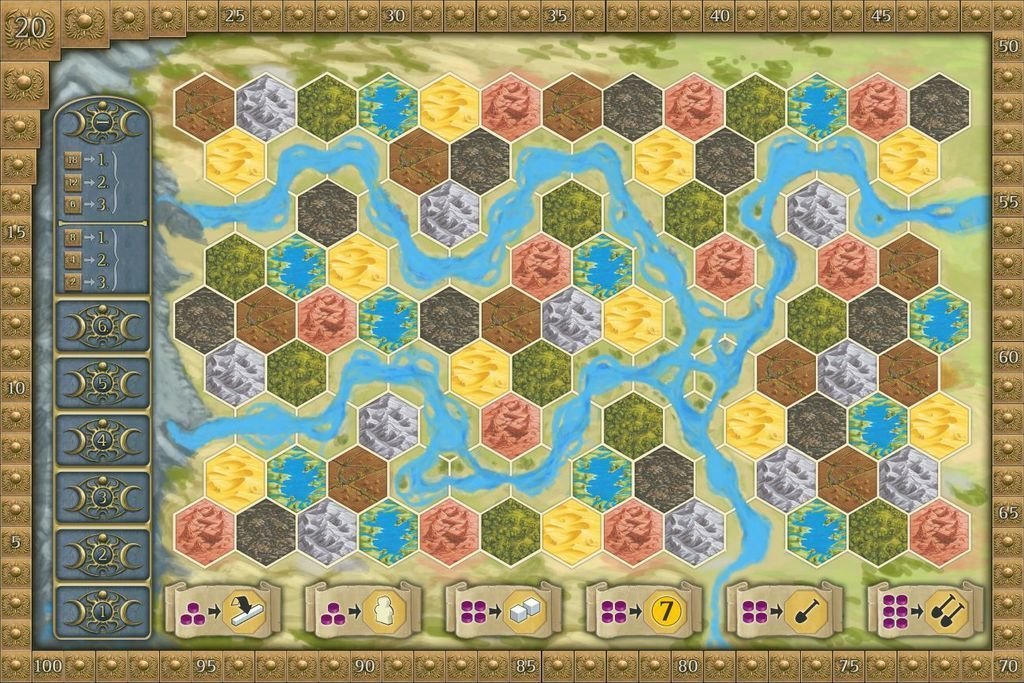}
    \caption{The Terra Mystica Game Board and Its Representation}
    \label{fig:TM_Board}
\end{figure}

We begin by noting that the TM GameBoard \ref{fig:TM_Board} is naturally represented as $13 \times 9$ hexagonal grid. As mentioned in the challenges section, this presents a unique problem since for each intuitive ``tile``, we have $6$ rather than the the $4$ (as in Go, Chess, and Shogi). Furthermore, unlike chess where a natural dilation of the convolution will cover additional tangent spots equally (expanding to $8$), the hexagonal nature makes TM particularly interesting.

However, a particular peculiarity of TM is that we can think of each ``row'' of tiles as being shifted by ``half'' a tile, thereby becoming ``neighbors''. With this approach, we chose to instead represent the TM board as a $9 \times 26$ grid, where each tile is horizontally doubled. Our terrain representation of the map then begins as a $9 \times 26 \times 8$ stack of layers. Each layer is a binary encoding of the terrain-type for each tile. The $7$ main types are \{PLAIN, SWAMP, LAKE, FOREST, MOUNTAIN, WASTELAND, DESERT \}. It as a possible action to ``terra-form`` any of these tiles into any of the other available tiles, therefore why we must maintain all $7$ of them as part of our configuration. The $8$-th layer actually remains constant throughout the game, as this layer represents the water-ways and cannot be modified. Note that even-row ($B, D, F, H$) are padded at columns $0$ and $25$ with $\text{WATER}$ tiles.

The next feature which we tackle is the representation of the structures which can be built on each terrain tile. As part of the rules of TM, a structure can only exists on a terrain which corresponds to it's player's terrain. As such, for each tile we only need to consider the $5$ possible structures, \{DWELLING, TRADING POST, SANCTUARY, TEMPLE, STRONGHOLD \}. We encode these as an additional five-layers in our grid. Our representation is now a $9 \times 26 \times 13$ stack.

We now proceed to add a set of constant layers. First, to represent each of the 5 special-actions, we add $6$-constant layers which will be either $0$ or $1$ signifying whether a particular action is available ($0$) or take ($1$). This gives us a $9 \times 26 \times 19$ representation.

To represent the scoring tiles (of which there are 8), we add $8\times6$ constant layers (either all $1$ or all $0$) indicating their presence in each of the $6$ rounds. This gives us a $9 \times 26 \times 75$ stack.

For favor tiles, there are 12 distinct favor tiles. We add $12$ layers each specifying the number of favor tiles remaining. This gives use $9 \times 26 \times 87$.

For the bonus tiles, we add $9$ constant layers. These $9$ layers specify which favor tiles were selected for this game (only $P + 3$ cards are ever in play). This gives us a game-board representation which is of size $9 \times 26 \times 96$

\subsubsection{Player Representation and Resource Limitations}
\label{subsubsection:player_representation_and_resource_limitations}
We now introduce another particularity of TM, which is the fact that \textbf{each} player has a different amount of \textbf{resources} which must be handled with care. This is something which is not treated in other games, since the resource limitation does not exist in Go, Chess, or Shogi (other than those fully encoded by the state of the board).

With that in-mind, we move to the task of encoding each player. To be fully generic, we scale this representation with the number of players playing the game, in our case, $P = 2$.

To do this, for each player, we add constant layers specifying: (1) number of workers, (2) number of priests,  (3) number of coins, (4) power in bowl I, (5) power in bowl II, (6) power in bowl III, (7) the cost to terraform, (8) shipping distance, (9-12) positions in each of the 4 cult tracks, (13-17) number of building built of each type, (18) current score, (19) next round worker income, (20) next round priest income, (21) next round coin income, (22) next round power income, (23) number of available bridges. This gives us a total of $23P$ additional layers required to specify information about the player resources.

Next, we consider representing the location of bridges. We add $P$ layers, each corresponding to each player, in a fixed order. The each layer is a bit representing the existence/absence of a bridge at a particular location. This gives us $24P$ layers.

We've already considered the positions of the player in the cult-track. The only thing left is the tiles which the player may have. We add $9 + 10 + 5$ layers to each player. The first $9$ specify which bonus card the player currently holds. The next $10$ specify which favor tiles the player currently owns. And the last $5$ specify how many town tiles of each type the player currently holds. This gives use an additional $24P$ layers.

We end with a complete stack of dimension $9 \times 26 \times 24P$ to represent $P$ players.

\subsubsection{Putting it All Together}
\label{subsubsection:final_input}
Finally, we add $14$ layers to specify which of the $14$ possible factions the neural network should play as. This gives us an input representation of size $9 \times 26 \times (48P + 110)$. See Table \ref{table:input_size_comparison} which places this into context. In our case, this becomes $9 \times 26 \times 206$.

\begin{table}[!ht]
\begin{tabular}{|l|l|l|}
\hline
\textit{\textbf{Domain}} & \textbf{Input Dimensions} & \textbf{Total Size} \\ \hline
\textbf{Atari 2600}      & 84 x 84 x 4               & 28,224              \\ \hline
\textbf{Go}              & 19 x 19 x 17              & 6,137               \\ \hline
\textbf{Chess}           & 8 x 8 x 119               & 7,616               \\ \hline
\textbf{Shogi}           & 9 x 9 x 362               & 29,322              \\ \hline
\textbf{Terra Mystica}   & 9 x 26 x 206              & 48,204              \\ \hline
\textbf{ImageNet}        & 224x224x1                 & 50,176              \\ \hline
\end{tabular}
\caption{Comparison of input sizes for different domain of both games. For reference, typical CNN domain of ImageNet is also included.}
\label{table:input_size_comparison}
\end{table}

\subsection{Action Space Representation}
\label{section:action_space_representation}
Terra Mystica is a complex game, where actions are significantly varied. In fact, it is not immediately obvious how to even represent all of the possible actions. We provide a brief overview here of our approach.

In general, there are 8 possible actions in TM which are, generally speaking, quite distinct. In general, we output all possible actions and assign a probability. Illegal actions are removed by setting their probabilities to zero and re-normalizing the remaining actions. Actions are considered legal as long as they can be legally performed during that turn (ie, a player can and will burn power/workers/etc. in order to perform the required action. We could technically add additional actios for each of this possibilities, but this vastly increases the complexity.

\begin{enumerate}
    \item Terra Form and Build: This action consists of (1) selecting a location (2) selecting a terrain to terraform to (if any) and (3) optionally selecting to build.  We can represent this process as a vector of size $9 \times 13 \times (7\times 2)$. The $9 \times 13$ is selecting a location, while the first $7$ layers the probability of terraforming into one of the $7$ terrains and \textbf{not} building, and the last $7$ the probability of terraforming into each of the $7$ terrains and building.
    \item Advancing on the Shipping Track: The player may choose to advance on the shipping track. This consists of a single additional value encoding the probability of advancement.
    \item: Lowering Spade Exchange Rate: The player may choose to advance on the spade track. This consists of a single additional value encoding the probability of choosing to advance on the spade track.
    \item Upgrading a Structure: This action consists of (1) selecting a location, (2) selecting which structure to upgrade to. Depending on the location and existing structure, some actions may be illegal. We can represent this process as a vector of size $9 \times 13 \times 4$ specifying which location as well as the requested upgrade to the structure (DWELLING to TRADING POST, TRADING POST to STRONG HOLD, TRADING POST to TEMPLE, or TEMPLE to SANCTUARY).
    
    Note that when a structure is built, it's possible for the opponents to trade victory points for power. While this is an interesting aspect of the game, we ignore for our purposes and assume players will never chose to take additional power.
    \item Send A Priest to the Order of A  Cult: In this action, the player choose to send his or her priest to one of four possible cults. Additionally, the player must determine if he wants to send his priest to advance $3,2$ or $1$ spaces -- some of which maybe illegal moves. We can represent this simply as a $4 \times 3$ vector of probabilities.
    \item Take a Board Power Action: There are 6 available power actions on the board. We represent this as a $6 \times 1$ vector indicating which power action the player wishes to take. Actions can only be take once per round.
    \item Take a Special Action: There are multiple possible ``special`` actions a player may choose to take. For example, there's a (1) spade bonus tile, (2) cult favor tile as well as special action allowed by the faction (3). As such, we output a $3 \times 1$ vector in this case for each of the above mentioned actions, many of which may be illegal. 
    \item Pass: The player may chose to pass. If the first to pass, the player will become the first to go next round. For this action, the player must also chose which bonus tile to take. There are $9$ possible bonus tiles (some which won't be available, either because they were never in play or because the other players have taken them). As such, we represent this action by a $9 \times 1$ vector. 
    \item Miscellaneous: At any point during game-play for this player, it may become the case that a town is founded. For each player round, we also output a $5 \times 1$ vector of probabilities specifying which town tile to take in the even this has occurred. These probabilities are normalized independently of the other actions, as they are not exclusive, though most of the time they will be ignored since towns are founded relatively rarely (two or three times per game).
\end{enumerate}

\subsubsection{Concluding Remarks for Action Space Representation}
\label{subsubsection:concluding_remarks_for_action_space_representation}
As described above, this leads to a relatively complex action-space representation. In fact, we'll end-up outputting a $9 \times 13 \times 18 + 4 \times 3 + 20 + 5 $ We summarize the action-space representation in Table \ref{table:output_size_comparison} and provide a few other methods fore reference.

\begin{table}[!ht]
\begin{tabular}{|l|l|l|}
\hline
\textit{\textbf{Domain}} & \textbf{Input Dimensions} & \textbf{Total Size} \\ \hline
\textbf{Atari 2600}      & 18 x 1               & 18              \\ \hline
\textbf{Go}              & 19 x 19 + 1              & 362               \\ \hline
\textbf{Chess}           & 8 x 8 x 73               & 4,672               \\ \hline
\textbf{Shogi}           & 9 x 9 x 139               & 11,259              \\ \hline
\textbf{Terra Mystica}   & 9 x 13 x 18 + 4x3 + 25 &   2,143            \\ \hline
\textbf{ImageNet}        & 1000x1                 & 1000              \\ \hline
\end{tabular}
\caption{Comparison of action space sizes for different domain of both games. For ImageNet, we consider the class-distribution the actions-space}
\label{table:output_size_comparison}
\end{table}

\subsection{Deep Q-Learning with MCTS Algorithm and Modifications}
\label{subsection:deep_qlearning_with_mcts_algorithm_and_modifications}
In this section, we present our main algorithm and the modifications we've made so-far to make it better suited for our state-space and action-space. We describe in detail the algorithm for completeness, despite the algorithm remaining mostly the same as that used in \cite{AlphaZero} and presented in detail in \cite{AlphaGoZero}.

\subsubsection{Major Modifications}
\label{subsubsection:major_modifications}
The main modifications to the algorithm are mostly performed on the architecture of the neural network. 
\begin{enumerate}
    \item We extend the concept of introduced in \cite{AlphaGoZero} of ``dual'' heads to ``multi-head`` thereby providing us with multiple differing final processing steps. 
    \item We modify the output and input dimensions accordingly.
\end{enumerate}

\subsubsection{Overview of Approach}
\label{subsubection:overview_of_approach}
We make use of a deep neural network (architecture described in Section \ref{subsubsection:neural_network_architecture}) $(\textbf{p}, \textbf{m}, v) = f_{\theta}(s)$ with parameters $\theta$, state-representations $s$, as described in Section \ref{subsection:input_representation}, output action-space $textbf{p}$, as described in Section \ref{section:action_space_representation}, additional action-information as $\textbf{m}$ as described in Section \ref{subsubsection:concluding_remarks_for_action_space_representation}, and a scalar value $v$ that estimates the expected output $z$ from position $s$ $v \approx \mathbb{E}[z \mid s]$. The values from this neural network are then used to guide MCTS, and the resulting move-probabilities and game output are used to iteratively update the weights for the network.

\subsubsection{Monte Carlo Tree Search} 
We provide a brief overview of the MCTS algorithm used. We assume some familiarity with how MCTS works. In general, there are three phases which need to be considered. For any given state $S_t$ which we call the root-state (this is the current state of gameplay), the algorithm simulates 800 iterations of gameplay. We note that AlphaGoZero \cite{AlphaGoZero} does 1600 iterations and AlphaZero does \cite{AlphaZero} also does 800.

\begin{figure}
    \centering
    \includegraphics[scale=0.09]{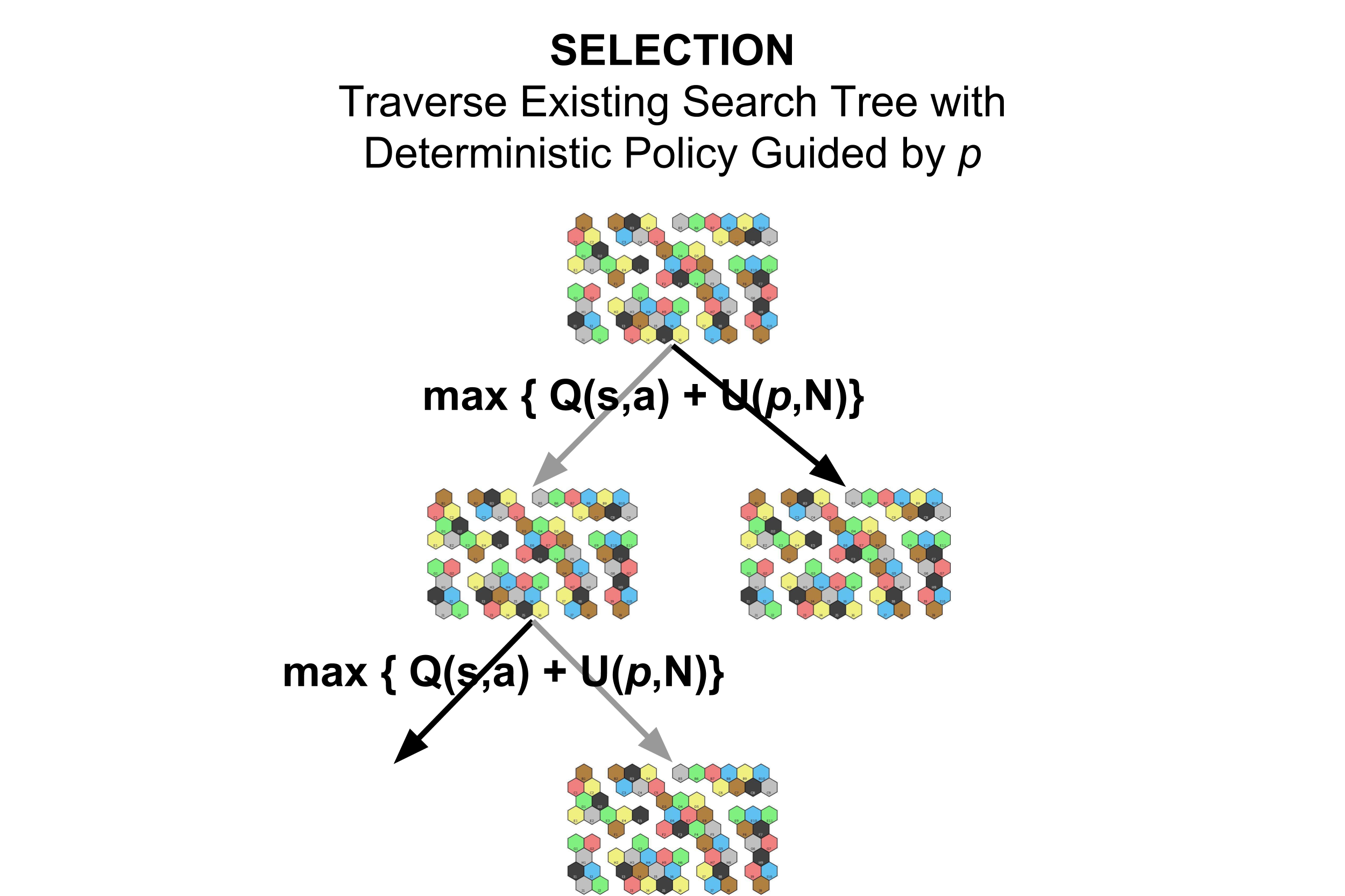}
    \caption{Monte Carlo Tree Search: Selection Phase. During this phase, starting from the root node, the algorithm selects the optimal path until reaching an un-expanded leaf node. In the case above, we select left action, then the right action, reaching a new, un-expanded board state.}
    \label{fig:mcts-selection}
\end{figure}

At each node, we perform a search until a leaf-node is found. A leaf-node is a game-state which has never-been encountered before. The search algorithm is relatively simple, as shown below and as illustrated in Figure \ref{fig:mcts-selection}. Note that the algorithm plays for the best possible move, with some bias given to moves with low-visit counts.

\begin{verbatim}
def Search(s):
    if IsEnd(s): return R
    if IsLeaf(s): return Expand(s)

    while seen(S):
        max_u, best_a = -INF, -1
        for a in Actions(S) :
            u = Q(s,a) + c*P(s,a)
                *sqrt(visit_count(s)))
                /(1+action_count(s,a))
            if u>max_u:
                max_u = u
                best_a = a
    s = Successor(s, best_a)
    v = search(sp, game, nnet)
\end{verbatim}

From above, we can see that as a sub-routing of search we have the `Expand` algorithm. The expansion algorithm is used to initialize non-terminal, unseen states of the game $S$ as follow.

\begin{figure}
    \centering
    \includegraphics[scale=0.09]{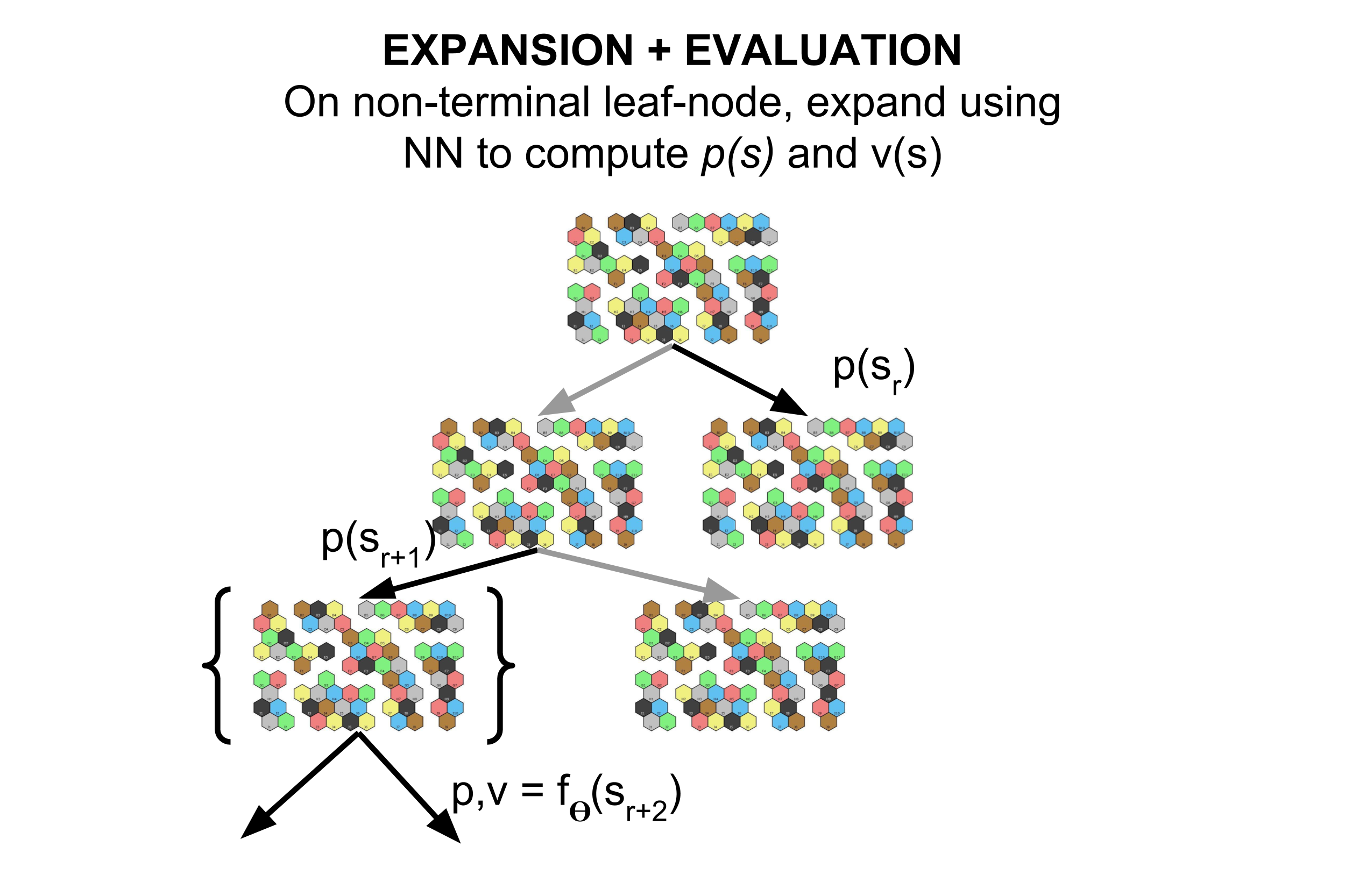}
    \caption{Monte Carlo Tree Search: Expansion Phase. During this phase, a leaf-node is ``expanded''. This is where the neural network comes in. At the leaf-node, we process the state $S_L$ to retrieve $p,v = f_{\theta}(S_L)$}, which is a vector of probabilities for the actions that are possible and a value function.
    \label{fig:mcts-expansion}
\end{figure}

\begin{verbatim}
def Expand(s):
    v, p = NN(s)
    InitializeCounts(s)
    StoreP(s, p)
    return v
\end{verbatim}
Where `InitializeCounts` simply initializes the counts for the new node (1 visit, 0 for each action). We also initialize all $Q(s,a) = 0$ and store the values predicted by our $NN$. Intuitively, we've now expanded the depth of our search tree, as illustrated in Figure \ref{fig:mcts-expansion}.

\begin{figure}
    \centering
    \includegraphics[scale=0.09]{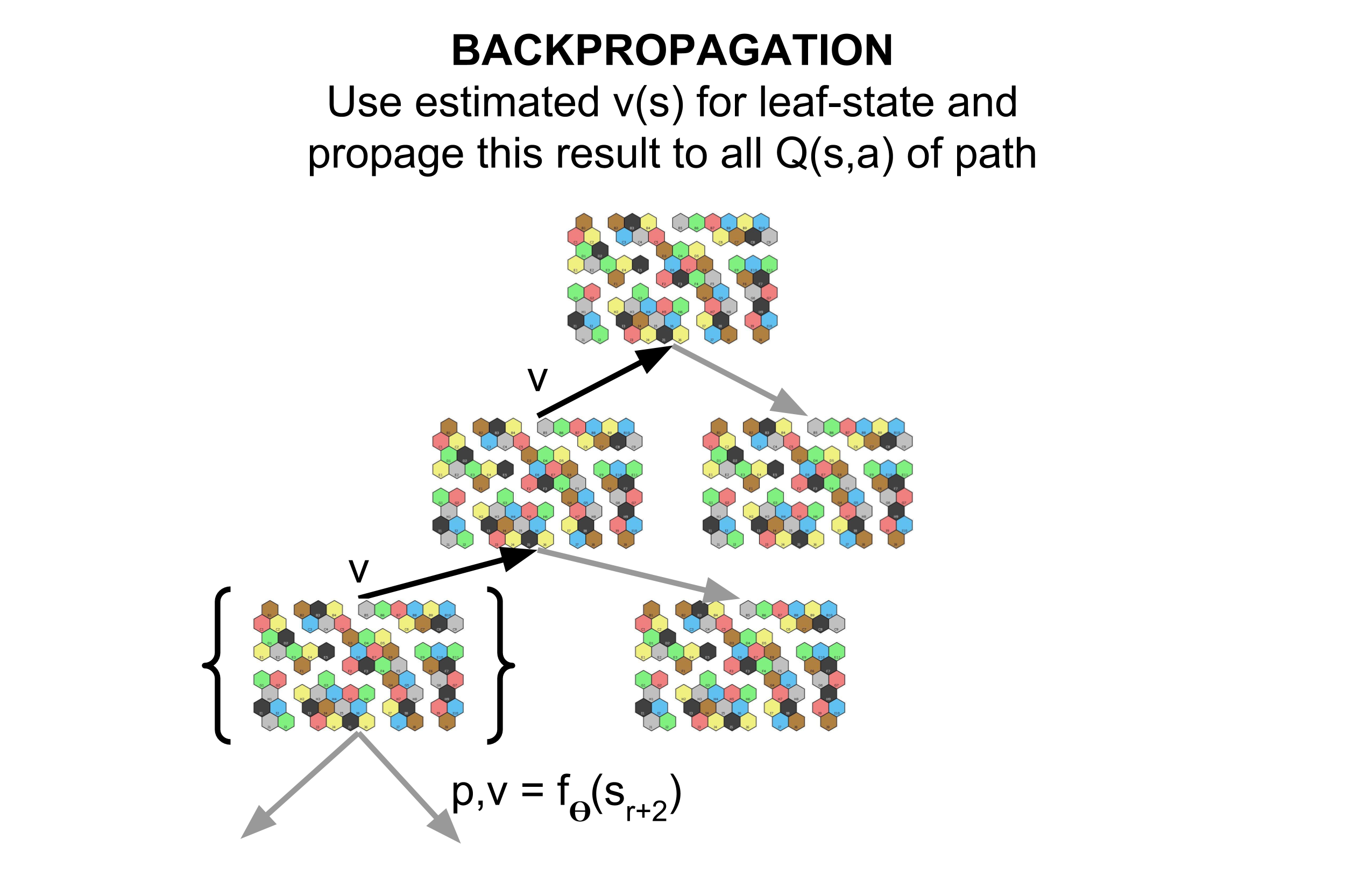}
    \caption{Monte Carlo Tree Search: Back-propagation Phase. During this phase, we use the value $v$ estimated at the leaf-node and propagate this information back-up the tree (along the path-taken) to update the stored $Q(s,a)$ values.}
    \label{fig:mcts-backpropagation}
\end{figure}

After the termination of the simulation (which ended either with an estimated $v$ by the NN or an actually $R$), we back-propagate the result by updating the corresponding $Q(s,a)$ values using the formula:
$$
Q(s,a) = V(Succ(s,a))
$$
This is outlined in Figure \ref{fig:mcts-backpropagation}.

\subsubsection{Neural Network Training}
The training can be summarized relatively straightforwardly. We batch $N$  (with $N = 128$) tuples $(s,p,v)$ and use this to train, with the loss presented in AlphaZero \cite{AlphaZero}. We use $c = 1$ to for our training. We perform back-propagation with this batch of data, and continue our games of self-play using the newly updated neural network. This is all performed synchronously.

\subsubsection{Neural Network Architecture}
\label{subsubsection:neural_network_architecture}
\begin{figure}[!ht]
    \centering
    \includegraphics[scale=0.09]{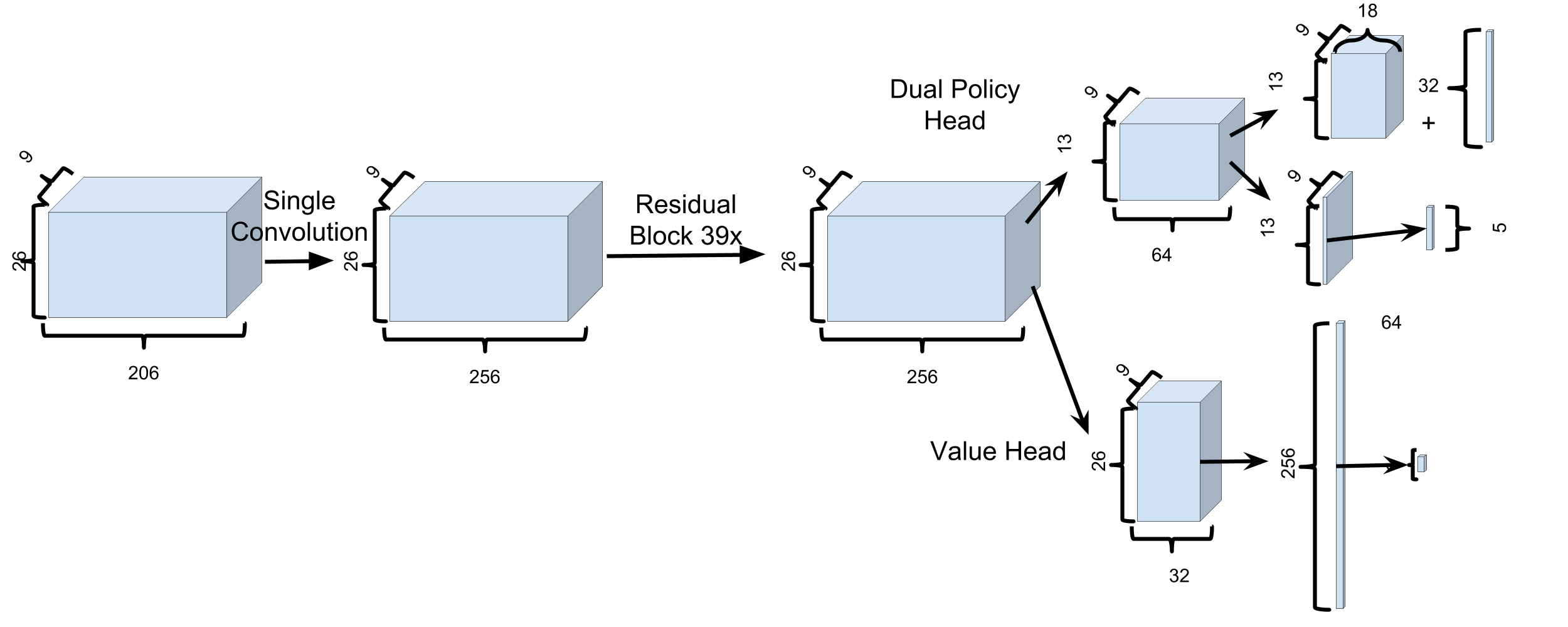}
    \caption{Detailed diagram of the multi-headed architecture explored for the game of Terra Mystica.}
    \label{fig:architecture}
\end{figure}

We first describe the architecture of our neural network. For brief overview, see Figure \ref{fig:architecture}. The input is a $9 \times 26 \times 206$ image stack as described in Section \ref{subsubsection:final_input}. Note that unlike other games, our representation includes no temporal-information ($T = 1$), both for computational efficiency and due to the fact that the game is fully encoded with the current state. 

The input features $s_t$ are processed by a residual tower which consists of a single convolution block  followed by $19$ residual blocks, as per \cite{AlphaGoZero}. The first convolution block consists of 256 filters of kernel size $3 \times 3$ with stride 1, batch normalization, and a ReLU non-linearity. 

\begin{figure}[!ht]
    \centering
    \includegraphics[scale=0.3]{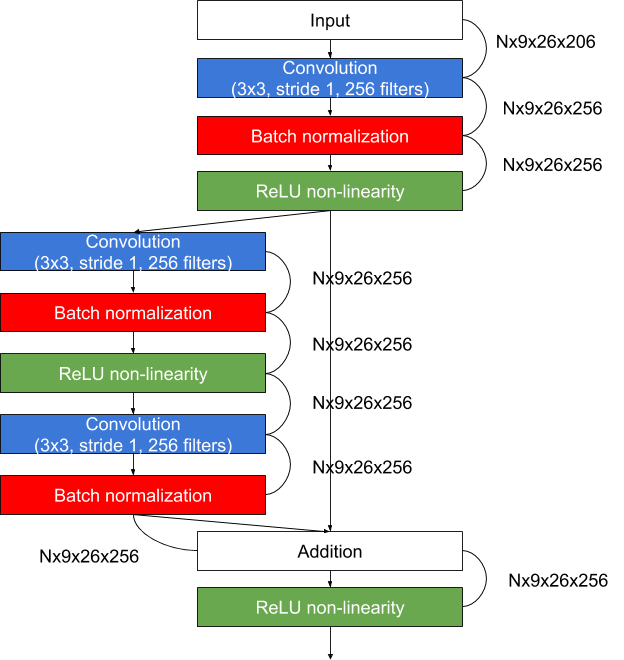}
    \caption{Architecture Diagram for shared processing of the state space features. An initial convolution block is used to standardize the number of features, which is then followed by 18 residual blocks.}
    \label{fig:input_architecture_diagram}
\end{figure}

Each residual block applies the following modules, in sequence, to its input. A convolution of 256 filters of kernel size $3 \times 3$ with stride 1, batch-normalization, and a ReLU non-linearity, followed by a second convolution of 256 filters of kernel size $3 \times 3$ with stride $1$, and batch-normalization. The input is then added to this, a ReLu applied, and the final output taken as input for the next block. See Figure \ref{fig:input_architecture_diagram} for a reference.

\begin{figure}[!ht]
    \centering
    \includegraphics[scale=0.09]{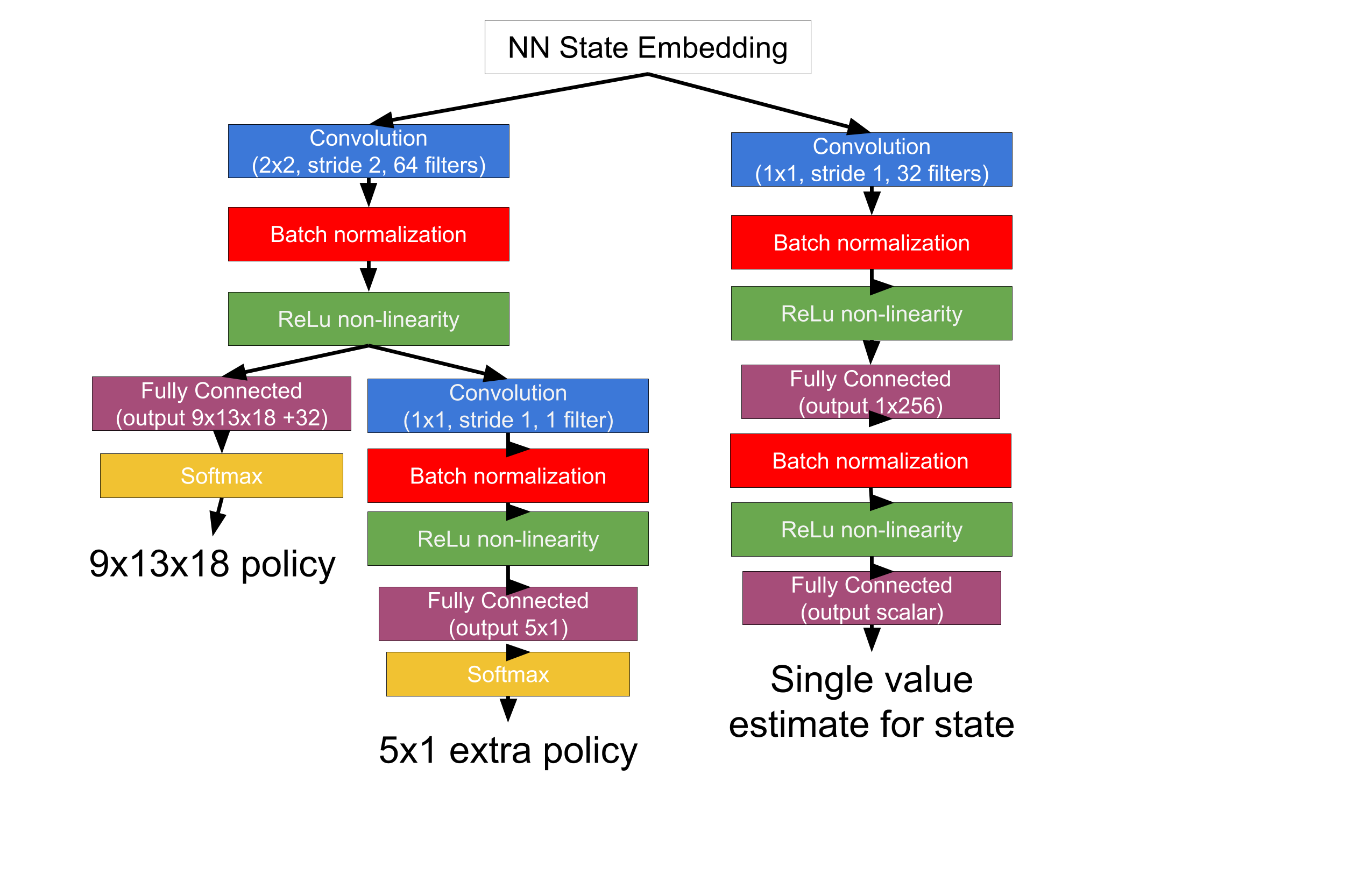}
    \caption{Details on multi-headed architecture for Neural Network. The final output state of the residual tower is fed into two paths. (1) On the left is the  policy network. (2) On the right is the value estimator. The policy network is further split into two, for computing two disjoint distributions over the action space, each normalized independently.}
    \label{fig:multi_head_architecture}
\end{figure}

The output of the residual tower is passed into multiple separate 'heads' for computing the policy, value, and miscellaneous information. The heads in charge of computing the policy apply the following modules, which we guess at given that the AlphaZero Paper \cite{AlphaZero} does not discuss in detail how the heads are modified to handle the final values. See Figure \ref{fig:multi_head_architecture} for an overview diagram.

For the policy, we have one head that applies a convolution of 64 filters with kernel size $2 \times 2$ with stride 2 along the horizontal direction, reducing our map to $9 \times 13 \times 64$. This convolution is followed by batch normalization, a ReLU non-linearity.

We then split this head into two further heads. We then apply an FC layer which outputs a vector of $9 \times 13 \times 18 + 32$ which we interpret as discussed in Section \ref{section:action_space_representation}, representing the mutually exclusive possible actions that a player can take.

For the second part, we apply a further convolution with 1 filter of size $1 \times 1$, reducing our input to $9 \times 13 \time 1$. This is followed by a batch-normalization layer followed by a ReLU. We then apply a FC layer further producing a probability distribution over a $5 \times 1$ vector.

For the value head, we apply a convolution of $32$ filters of kernel size $1 \times 1$ with stride $1$ to the, followed by batch normalization and a ReLU unit. We follow this with an FC to 264 units, a ReLU, another FC to a scalar, and a tanh-nonlinearity to output a scalar in the range $[-1, 1]$.

\section{Experimental Results}
Given the complexity of the problem we're facing, the majority of the work has been spent developing the reinforcement learning pipeline with an implementation of MCTS. The pipeline appears to not train well, even after multiple games of self-play.

\subsection{Baselines}
For the base-lines, we compare the final scores achieved by existing AI agents. We see their results in Table \ref{table:average_2p_score}. The results demonstrate that current AIs are fairly capable at scoring highly during games of self-play.

\begin{table}[!ht]
\begin{tabular}{|l|l|l|}
\hline
\multicolumn{3}{|c|}{{\ul \textbf{Simulated Self-Play Average Scores - AI}}}      \\ \hline
\textbf{Faction} & \textbf{Average Score} & \textbf{Sampled Games} \\ \hline
Halfing          & 92.21                 & 1000                   \\ \hline
Engineers        & 77.12                 & 1000                   \\ \hline
\end{tabular}
\caption{Self-play easy AI agent: AI\_Level5 from \cite{TMStatsAI}}
\label{table:average_2p_score_ai}
\end{table}

\subsection{Oracle}
A second comparison, showing in Table \ref{table:average_2p_score}, demonstrates the skill we would expect to achieve. These are the average scores of the best human players, averaged over online data.

\begin{table}[!ht]
\begin{tabular}{|l|l|l|}
\hline
\multicolumn{3}{|c|}{{\ul \textbf{Average Human Score (2p)}}}      \\ \hline
\textbf{Faction} & \textbf{Average Score} & \textbf{Sampled Games} \\ \hline
Halfling          & 133.32                 & 2227                   \\ \hline
Engineers        & 127.72                 & 1543                   \\ \hline
\end{tabular}
\caption{Average human scores by faction for a two-player TM games online.}
\label{table:average_2p_score}
\end{table}

\subsection{AlphaTM}
The results for AlphaTM are presented below. Training appears to not have taken place, at least not with the architecture and number of iterations which we executed. The AI still appears to play randomly, specially at later, and more crucial, stages of the game. See Table \ref{table:average_2p_score_alpha_tm}.

\begin{table}[!ht]
\begin{tabular}{|l|l|l|l|}
\hline
\multicolumn{3}{|c|}{{\ul \textbf{Simulated Self-Play Average Scores - AlphaTM}}}      \\ \hline
\textbf{Faction} & \textbf{Average Score} & \textbf{Training Iterations} \\ \hline
Halfing          & 32.11                 & 10,000                   \\ \hline
Engineers        & 34.12                 & 10,000                 \\ \hline
\end{tabular}
\caption{Our Self-Play AI after 10,000 training iterations, with average score over final 1000 games.}
\label{table:average_2p_score_alpha_tm}
\end{table}

Overall, we summarize:
\begin{itemize}
    \item The AI plays poorly in early stages of the game, though it seems to learn to build structures adjacent to other players.
    \item As the game progresses, the actions of the agent are indistinguishable from random. A cursory analysis of $\pi$ reveals these states are basically random. It appears that the AI is not learning to generalize, or has simple not played sufficient games.
\end{itemize}

\section{Future Work}
Given the poor results from the experiments above, many avenues exists for future work. In particular, we propose a few extensions to the above approach below.

\subsection{Generalizing to Multi-Player Game}
In the general context, the reinforcement learning pipeline that performed the best (with some semblance of learning) is the one where the game was presented as a zero-sum two-player game explicitly (I win, you lose). While the neural network architecture presented can readily generalize to more players, the theory behind the learning algorithm will no-longer hold. The game is no longer zero-sum.

\subsection{New Architectures and Improvements} 
Another area of future work is experimenting with further architectures and general improvement, with possible hyper parameter tuning.

\section{Appendices}

\begin{figure}[!ht]
    \centering
    \includegraphics[scale=0.8]{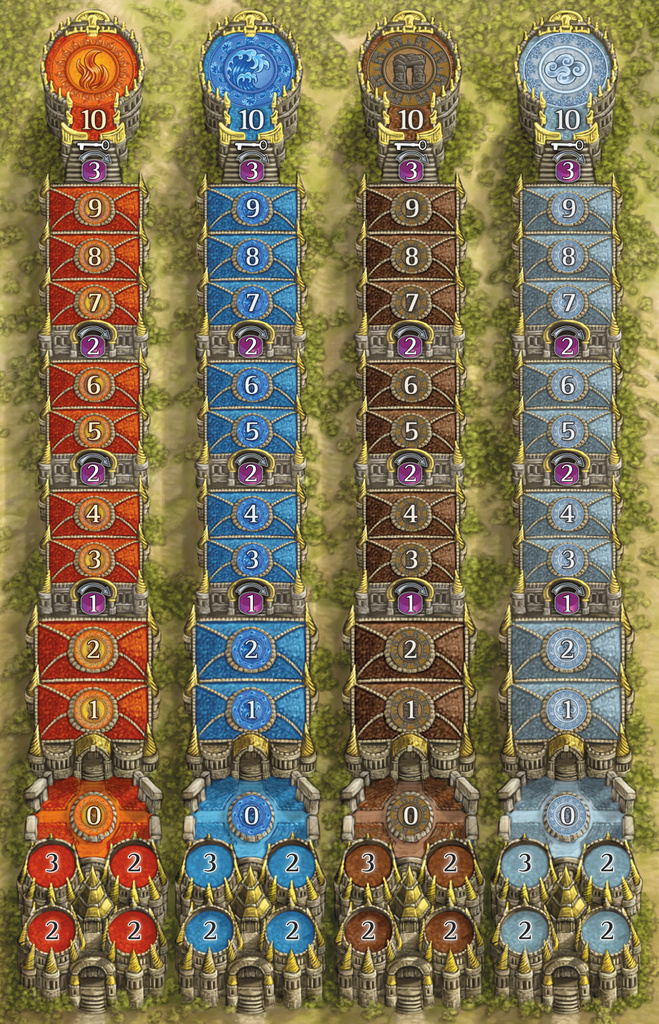}
    \caption{The Terra Mystica Cult Track}
    \label{fig:TM_Cult_Track}
\end{figure}

\begin{figure}[!ht]
    \centering
    \includegraphics[scale=0.20]{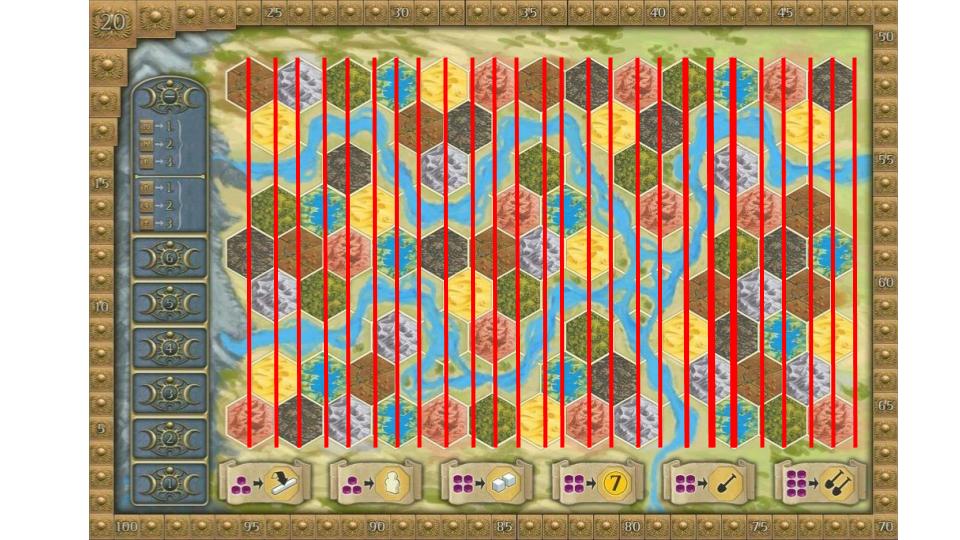}
    \caption{The Terra Mystica Board Representation}
    \label{fig:TM_Board_Representation}
\end{figure}

{\small
\bibliographystyle{common/ieee}
\bibliography{common/egbib}
}

\end{document}